\documentclass[aps,twocolumn,superscriptaddress,showpacs,floatfix]{revtex4-1}
\usepackage{amsfonts,amssymb,latexsym,xspace,epsfig,graphicx,color}
\usepackage{amsmath,enumerate,stmaryrd,xy,stackrel}

\usepackage[latin9]{inputenc}
\usepackage[T1]{fontenc}
\usepackage[english]{babel}
\usepackage{amsmath}
\usepackage[inner=2.5cm,outer=2.5cm,top=0.5cm,bottom=1.6cm,includeheadfoot]{geometry}
\usepackage{graphicx}
\usepackage{subfigure}
\usepackage{cancel}
\usepackage{ulem}
\usepackage{amssymb}
\usepackage{epsfig}
\usepackage{appendix}
\usepackage{graphicx}
\usepackage{bm}
\usepackage{dcolumn}
\usepackage{multirow}
\usepackage{float}
\usepackage{listings}

\include{srctex}

\newcommand{\pard}[2][]{\frac{\partial#1}{\partial#2}}

\newcommand{\phidnz}{\phi_{\rm dn}}
\newcommand{\phidno}{\phi_{\rm\tilde{dn}}}
\newcommand{\phidc}{\phi_{\rm dc}}

\begin{document}

\title{Bose-Einstein condensate confined in a 1D ring stirred with a rotating delta link}

\author{Axel P\'erez-Obiol}
\affiliation{Laboratory of Physics, Kochi University of Technology, Tosa Yamada, Kochi 782-8502, Japan}
\author{Taksu Cheon}
\affiliation{Laboratory of Physics, Kochi University of Technology, Tosa Yamada, Kochi 782-8502, Japan}

\begin{abstract}
We consider a Bose-Einstein condensate with repulsive interactions confined in a 1D ring where a Dirac delta is rotating at constant speed.
The spectrum of stationary solutions in the delta comoving frame is analyzed in terms of the nonlinear coupling,
delta velocity, and delta strength, which may take positive and negative values.
It is organized into a set of energy levels conforming a multiple swallowtail structure
in parameter space, consisting in bright solitons, gray and dark solitonic trains, and vortex states.
Analytical expressions in terms of Jacobi elliptic functions
are provided for the wave functions and chemical potentials.
We compute the critical velocities and perform a Bogoliubov analysis for the ground state
and first few excited levels, establishing possible adiabatic transitions between the stationary and stable solutions.
A set of adiabatic cycles is proposed in which gray and dark solitons, and vortex states of arbitrary quantized angular momenta, are obtained
from the ground state by setting and unsetting a rotating delta.
These cycles are reproduced by simulations of the time-dependent Gross-Pitaevskii equation
with a rotating Gaussian link.
\end{abstract}

\date{\today}

\maketitle

\section{Introduction}

Bose-Einstein condensates (BECs) constrained in annular traps provide a way to study
various phenomena related to superfluidity, including persistent currents
and their decay, phase slips, and critical velocity \cite{ryu07,moulder12,beattie13,wright13pra}.
Persistent currents can be created experimentally by the
 application of artificial gauge fields \cite{dalibard11}, or by rotating
a localized, tunable repulsive barrier around the ring
\cite{ramanathan11,piazza09,piazza13,wright13prl}.
Through the latter method, 
 hysteresis between different circulation states was
observed in \cite{eckel14}.
Within the context of atomtronics, a {\it ring condensate}
is a key atomic circuit element. It has demonstrated
its capability as a superconducting quantum interference device \cite{ryu13},
entailing the possibility of high precision measurements 
and applications in quantum information processing \cite{hallwood10,schenke11,amico14}.

In the view of a better control of BECs, phase transitions
have been analyzed in different ring settings within the mean field approach.
They were first studied in a ring under a rotational drive \cite{kanamoto09},
and then through the interplay between rotation and symmetry breaking potentials 
or rotating lattice rings.
One lattice site was studied in \cite{fialko12}, a double well in \cite{li12},
and a more general unified approach of a ring lattice in \cite{munoz19},
all involving the possibility to adiabatically connect
different quantized states such as persistent currents or solitons.

By solving the Gross-Pitaevski equation (GPE), various works have studied
the energy diagram and metastability of BECs in rings with a rotating defect
\cite{baharian13,munoz15,kunimi18}.
In the case of the 1D GPE, stationary solutions can be found through the inverse scattering method
 or by directly integrating and writing them in terms of Jacobi functions.
These solutions have been analyzed under box and periodic boundary conditions \cite{carr002},
under a rotational drive \cite{kanamoto09}, and under some specific constant potentials \cite{seaman05}.
The flow past an obstacle in the form of a Dirac delta was studied perturbatively in 
\cite{hakim97,pavloff02}. In \cite{cominotti14} a 1D ring with a
rotating Dirac delta was analyzed for some specific rotations,
strengths, and nonlinearities.

In this paper, we study a repulsive BEC in a 1D ring where a Dirac delta link is
 rotating at constant speed.
The use of analytical solutions, expressed in terms of Jacobi functions,
 allows us to compute the stationary wave functions and chemical potentials
for the ground state and an arbitrary number of excited energy levels.
The obtained energy diagram, depending on the delta velocities and strengths,
both attractive and repulsive, is analyzed as a function of the coupling strength.
This diagram entails a series of critical velocities which,
together with a Bogoliubov analysis, lay out the distribution 
of stable and metastable states in parameter space,
and which adiabatic transitions between them are possible.
Within these transitions, we propose a few adiabatic cycles
in which excited solitonic and vortex states are produced by setting and unsetting a rotating
delta.

This paper is organized as follows. 
In the next section, \ref{sec:model}, we introduce
the theoretical model, defining the GPE and boundary conditions
in the Dirac delta comoving frame,
and provide a method to compute the spectrum.
The results are in Sec.~\ref{sec:results}, in which we illustrate
the main features of the spectrum (\ref{sec:spectrum}),  its stability (\ref{sec:metastability}), 
and its dependence on the nonlinearity (\ref{sec:g}).
In Sec.~\ref{sec:excitation}, we propose a set of adiabatic 
paths to excite the condensate. We conclude this paper in Sec.~\ref{sec:conclusions}.
Mathematical details are found in the Appendices.

\section{Theoretical model}
\label{sec:model}

We consider a BEC at zero temperature in a tightly transverse annular trap
in which a Dirac delta link is rotating at constant speed. The point-like potential
is chosen instead of a finite one such that analytical solutions can be obtained,
with the view that the results may not qualitatively change with respect
to a very peaked Gaussian.
Considering only stationary solutions, and within the mean field approach,
we can determine the condensate wave function in the delta comoving frame, $\phi(\theta)$,
by the 1D Gross-Pitaevskii equation. Then, $\phi(\theta)$ is constrained
by delta boundary conditions and normalization,
\begin{align}
\label{eq:gpf}
-\frac12 \phi''(\theta)+g|\phi(\theta)|^2\phi(\theta)=&\mu\,\phi(\theta),
\\
\label{eq:bc1}
\phi(0)-e^{i 2\pi\Omega}\phi(2\pi)=&0,
\\
\label{eq:bc2}
\phi'(0)-e^{i2\pi\Omega}\phi'(2\pi)=&\alpha\,\phi(0),
\\
\label{eq:norm}
\int_0^{2\pi}d\theta|\phi(\theta)|^2=&1,
\end{align}
where $g>0$ is the reduced 1D coupling, $\mu$ the chemical potential,
$\theta\in[0,2\pi)$, and $\Omega$ and $\frac{\alpha}{2}$ the velocity and strength of the delta link,
see App~\ref{sec:gpe}.
Here and in the rest of the paper we use units $\hbar=R=M=1$, $R$ being the radius
of the ring and $M$ the mass of the atoms.
Renormalizing a wave function $\phi(\theta)\to \sqrt{N}\phi(\theta)$
amounts to a rescaling of $g\to g\,N$. We choose to fix the normalization
and study how the spectrum depends on $g$.

Any solution of Eq.~(\ref{eq:gpf}), $\phi(\theta)=r(\theta)e^{i\,\beta(\theta)}$, can be written in closed form
in terms of a Jacobi elliptic function~\cite{seaman05}.
In particular, the density $\rho(\theta)\equiv r(\theta)^2$ depends linearly on the
square of one of the twelve Jacobi functions ($J$),
and the phase $\beta(\theta)$ is fixed by the density,
\begin{align}
\label{eq:density}
r^2_J(\theta)=&A+B\,J^2(k(\theta-\theta_j),m),
\\
\beta_J(\theta)=&\int_0^{\theta}d\tilde{\theta}\frac{\gamma}{r^2_J(\tilde{\theta})},
\end{align}
where $k$ is the frequency and $m\in(0,1)$ the elliptic
modulus, which generalizes the trigonometric and hyperbolic
functions into the Jacobi ones.
The constants $A$ and $B$, the shift $\theta_j$, $\gamma=\rho(\theta)\beta'(\theta)$,
a constant representing the current, and $\alpha$, $\Omega$ and $\mu$,
are fixed by Eqs.~(\ref{eq:gpf})-(\ref{eq:norm}) in terms of $k$ and $m$.
The spectrum $\mu(\alpha,\Omega)$ is thus given in parametric form,
$(\alpha(k,m),\Omega(k,m),\mu(k,m))$. Running $k$ and $m$ in a systematic
way in the three possible solutions allowed by Eqs.~(\ref{eq:gpf})-(\ref{eq:norm}),
 dn, $\tilde{{\rm dn}}$, and dc (see App.~\ref{sec:solutions})
we obtain $\mu(\alpha,\Omega)$ as a series of surfaces which fold onto each
other ---energy levels which cross and are degenerate at specific lines
$\Omega_{cr}(\alpha)$.
Any solution found for a specific $\alpha$ and $\Omega$,
also satisfies Eqs.~(\ref{eq:gpf})-(\ref{eq:norm})
with $\Omega\to\Omega\pm integer$ and $\beta(\theta)\to\pm\beta(\theta)$.
To obtain the complete spectrum, we shift and mirror the obtained spectrum
to $\Omega\to\Omega\pm integer$.

\section{Static properties}
\label{sec:results}

Our goal is to analyze the possible stable and adiabatic changes
of the condensate as one varies 
the strength and velocity of the Dirac delta.
For this we first study the structure of the spectrum $\mu(\alpha,\Omega)$, i.e.
 the regions in the $\alpha-\Omega$ plane in which stationary
solutions exist for the ground and first excited states,
and how the chemical potential depends on $\alpha$ and $\Omega$
(Sec.~\ref{sec:spectrum}).
Then we analyze whether the solutions at each region are
stable or metastable against a perturbation through a Bogoliubov analysis (Sec.~\ref{sec:metastability}).
The results in Sec.~\ref{sec:spectrum} and \ref{sec:metastability} are analyzed
and illustrated for $g=10$. In Sec.~\ref{sec:g} we study
how they depend on $g$.

\subsection{Spectrum}
\label{sec:spectrum}

\begin{figure}[t]
\centering
\includegraphics[width=.49\textwidth]{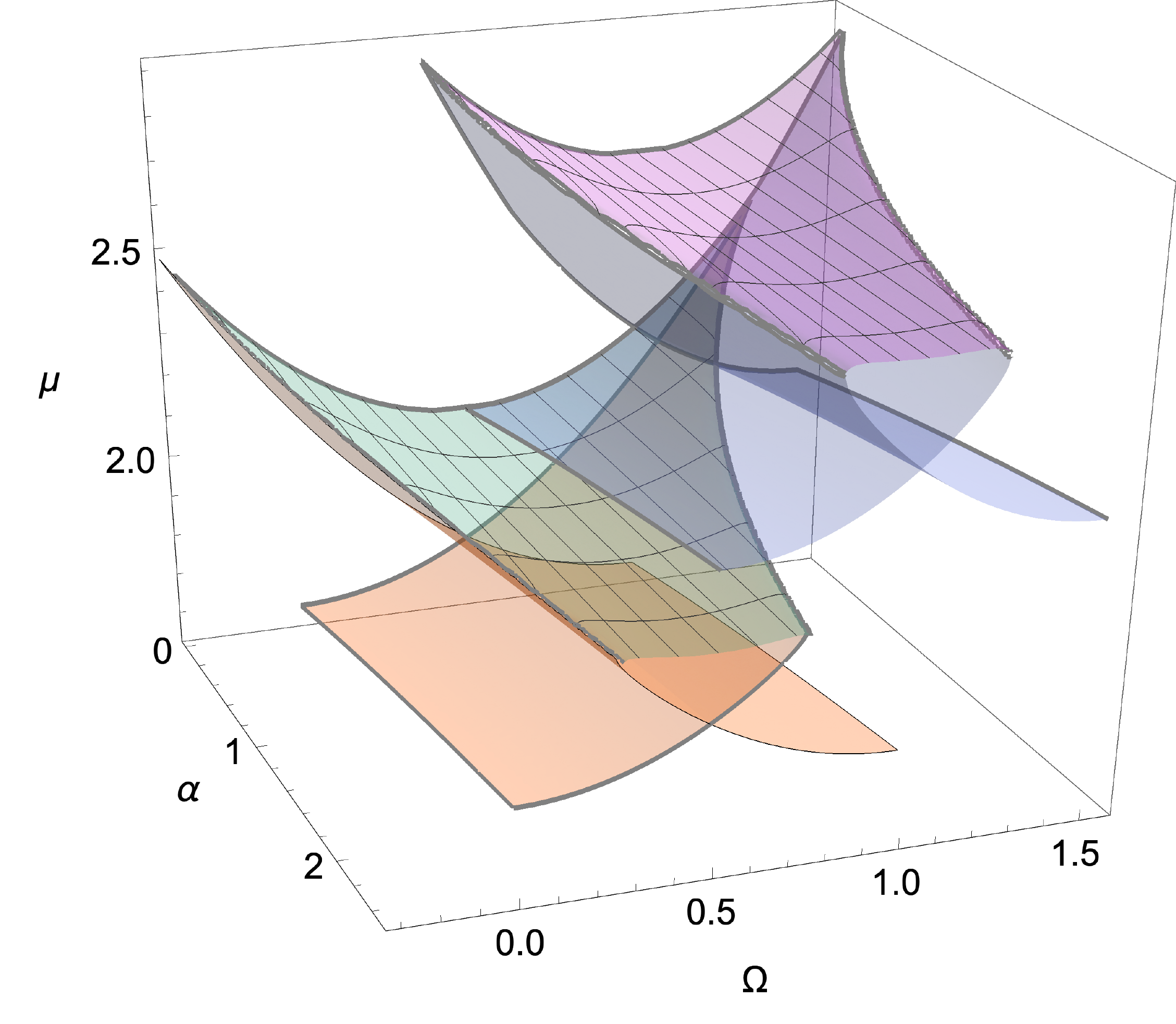}
\caption{(Color online)
Sample of the spectrum $\mu(\alpha,\Omega)$ for $\alpha>0$.
Each colored surface represents a set of solutions adiabatically
connected through variations of the delta link strength $\frac{\alpha}{2}$ and velocity
$\Omega$. Due to rotational symmetry, this structure can be shifted
$\Omega\to\Omega+integer$.
The spectrum at $\alpha<0$ is, qualitatively, a mirror image of the one at $\alpha>0$,
the surfaces being continuous (but not smooth) at $\alpha=0$.
The bottom 3D swallowtail structure (solid
red and green with grid levels) at $\alpha<0$ is an exception,
and its more complex structure is analyzed through its projections in Fig.~\ref{fig:regions}
at App.~\ref{sec:solutions}.
}
\label{fig:spectrum3d}
\end{figure}

\begin{figure*}[t]
\centering
\includegraphics[width=1\textwidth]{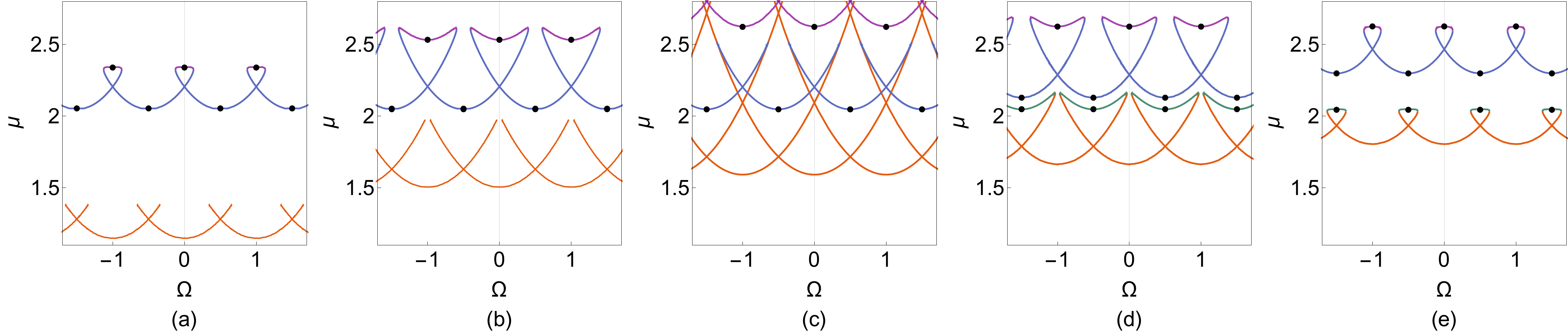}
\caption{(Color online). Sections $\alpha=-4,-1,0,1,4$ of the spectrum $\mu(\alpha,\Omega)$ for $g=10$ conforming a set of swallowtail diagrams. The parts in each diagram are colored
in correspondence to the regions in Figs.~\ref{fig:spectrum3d}
 and~\ref{fig:regions} to which they belong,
except the middle panel in which the swallowtails are not separated.
Black dots indicate the velocities and chemical potentials of dark solitonic trains.}
\label{fig:sections}
\end{figure*}

The spectrum $\mu(\alpha,\Omega)$ consists in a set of surfaces which cross
and merge at certain boundaries, as the ones plotted in Fig.~\ref{fig:spectrum3d}
and their symmetric versions at $\Omega\to\Omega+integer$.
Five sections of $\mu(\alpha,\Omega)$ with constant $\alpha$
are plotted in Fig.~\ref{fig:sections}.
Except for the ground state at $\alpha<0$,
they present a set of concatenated swallowtail (ST) shapes.
For any given $\alpha$, each series of swallowtail diagrams represents a set of stationary
solutions continuously connected among them through the parameters $k$ and $m$,
or through the velocity $\Omega$ and chemical potential $\mu$.
These type of energy diagrams are characteristic of hysteresis and
were analyzed in the context of ring condensates in \cite{mueller02}.
We organize and label them, and the surfaces that they constitute,
according to their position, ordered from lower to higher energy.
For each diagram we distinguish a bottom and a top part, both merging
at the tip of the swallowtail.
In the following we present their general features.
The structure of the spectrum is analyzed more thoroughly in App.~\ref{sec:solutions}.

Each energy level ---top or bottom part of a swallowtail diagram--- is symmetric with respect to $\Omega=\frac{l}{2}$ and
bounded by a pair of critical velocities $\frac{l}{2}\pm\Omega_{cr}(\alpha)$,
with $l$ an integer.
Solutions corresponding to a level
centered at a link velocity $\Omega=\frac{l}{2}$, are a boost from those
in the analogous level at $\Omega=0$ or $\Omega=\frac{1}{2}$.
The $n$th set of swallowtail diagrams, $n=1$ being the bottom one,
entails densities with $n$ depressions,
considering both the valleys characteristic of the Jacobi functions,
and the downward kinks in the case of $\alpha>0$.
We distinguish three types of solutions, depending on the depth of the depressions:
vortex states, dark solitons, and gray solitons.

{\it Vortex states}. The red parabolas of plot (c) ($\alpha=0$) in Fig.~\ref{fig:sections} centered at $l$
represent the chemical potential of vortex states of angular momentum $l$ as observed from the frame moving at $\Omega$. The minima of $\mu(\Omega)$ are at $\Omega=l$, where the observer is comoving with the vortex.

{\it Dark solitons}.
At precisely $\Omega=\frac{l}{2}$,
 solutions consist in dark solitonic trains (except for the ground state),
see the black dots in Fig.~\ref{fig:sections}.
In the delta comoving frame, the dark solitonic trains are stationary,
with zero current and constant phase, except for a phase jump of $\pi$ at each zero in the density.
They correspond to the minima
of the energy spectrum $\mu(\Omega)$ for any particular fixed $\alpha$.
In the lab frame, they comove with the condensate and the delta link at $\Omega=\frac{l}{2}$.

{\it Gray solitons}.
Solutions with velocities that depart from $\Omega=\frac{l}{2}$,
consist in gray solitonic trains, with shallower waves and faster currents
the larger $|\Omega-\frac{l}{2}|$.
At $\alpha=0$, gray solitonic trains with $n$ depressions
become completely flat and merge with vortex states at $\Omega=\frac{l}{2}\pm|\tilde{\Omega}_n-\frac{n}{2}|$, with
\begin{align}
\label{eq:omcr0}
\tilde{\Omega}_n=\sqrt{\frac{g}{2\pi}+\frac{n^2}{4}},
\end{align}
see  App.~\ref{sec:solutions} or \cite{carr002}.
At $\alpha\neq0$, the rotational symmetry is broken, and a pair of critical
velocities, corresponding to the tips of the swallowtails, 
limit the range of $\Omega$ for which stationary solutions exist.
In particular, the width of the bottom part of the $n$th swallowtail centered at $\frac{l}{2}$
monotonously decreases from $\frac{l}{2}\pm|\tilde{\Omega}_n-\frac{n-1}{2}|$, at $\alpha=0$,
 to $\frac{l}{2}\pm\frac{1}{2}$ at $|\alpha|\to\infty$.
The condensate in the ring is therefore able to sustain stationary solutions,
consisting in gray solitonic trains comoving with the kink, up
to a certain stirring velocity ---relative to $\frac{l}{2}$---,
which decreases with the strength of the delta link.

\begin{figure}[t]
\centering
\includegraphics[width=.48\textwidth,height=300pt]{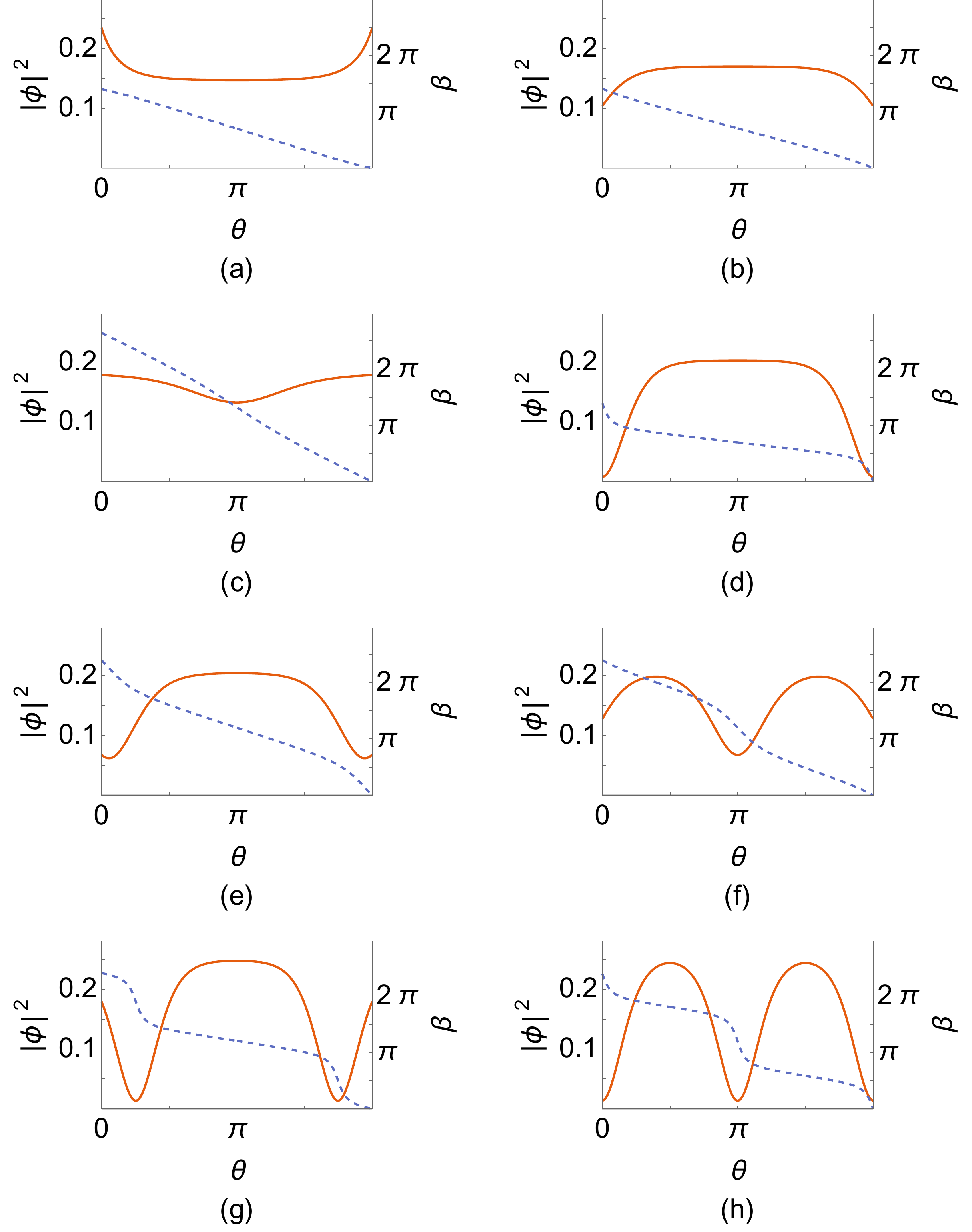}
\caption{(Color online). Densities (solid lines) and phases
(dashed lines) of eigenfunctions in the comoving frame characteristic of the eight
regions defined by the two first swallowtail diagrams.
The rows correspond, in order, to bottom and top of the first swallowtail diagram,
and bottom and top of the second one.
Left column plots correspond to $\alpha<0$ and the ones on the right
to $\alpha>0$.
All eigenfunctions are computed for $g=10$, and the specific
values of ($\alpha$, $\Omega$) are
(a) $(-1,0.7)$, (b) $(1,0.7)$,
(c) $(-0.02,1.32)$, (d) $(1,0.7)$,
(e) $(-1,1.2)$, (f) $(1,1.2)$,
(g) $(-1,1.2)$, (h) $(1,1.2)$.
}
\label{fig:eigenfunctions}
\end{figure}

A sample density and phase for each colored region
of Fig.~\ref{fig:spectrum3d} (and the corresponding ones
at $\alpha<0$),
are plotted in Fig.~\ref{fig:eigenfunctions}.
The densities corresponding to the bottom of first swallowtail surface at $\alpha<0$
entail an upward kink. This kink becomes higher and more peaked as the delta potential
becomes more attractive, and can be understood as a bright soliton.
In contrast, for the rest of levels, as $\alpha\to\pm\infty$,
the density at the delta position becomes zero. 
The densities corresponding to the first swallowtail diagram have one depression,
which for $\alpha<0$ consists of one valley and for $\alpha>0$ a downward kink.
Similarly, the four plots corresponding to the second swallowtail levels
have two depressions. For $\alpha>0$, one of these depressions is also understood
as the downward kink imposed by the delta. From these plots it can also be inferred
the relation between the the phase and the density, $\beta'(\theta)=\frac{\gamma}{\rho(\theta)}$,
which implies higher phase gradients (velocities) for lower densities.

\subsection{Metastability}
\label{sec:metastability}

A Dirac delta with fixed strength and rotating at a constant speed allows
 an infinite set of solutions organized in chemical potential levels.
The stability of these solutions can be studied by adding a small perturbation
to the stationary wave function
\begin{align}
\label{eq:pert}
\Psi=e^{-i\mu t}(\phi+u\,e^{-i\omega t}-v^*e^{i\omega^* t}), 
\end{align}
and analyzing how it evolves.
Replacing this function in the time-dependent Gross-Pitaevskii equation (Eq.~(\ref{eq:gp2}))
and linearizing in $u$ and $v$, we obtain the corresponding
Bogoliubov system of equations \cite{bogoliubov47},
\begin{align}
\label{eq:bog1}
-\frac12u''+2g\,|\phi|^2u-\mu\,u-g\phi^2v=w\,u,
\\
\label{eq:bog2}
\frac12v''-2g\,|\phi|^2v+\mu\,v+g{\phi^*}^2u=w\,v.
\end{align}
Both perturbations, $u$ and $v$, must satisfy the boundary conditions separately,
which read,
\begin{align}
u(0)-e^{i 2\pi\Omega}u(2\pi)=&0,
\\
u'(0)-e^{i 2\pi\Omega}u'(2\pi)=&\alpha\,u(0),
\\
v(0)-e^{-i 2\pi\Omega}v(2\pi)=&0,
\\
v'(0)-e^{-i 2\pi\Omega}v'(2\pi)=&\alpha\,v(0).
\end{align}
We solve this system of equations, by changing variables to 
\begin{align}
u(\theta)=e^{-i\,\Omega \,\theta}\tilde{u}(\theta),
\\
v(\theta)=e^{i\,\Omega\, \theta}\tilde{v}(\theta),
\end{align}
and then expanding $\tilde{u}$ and $\tilde{v}$
in an orthonormal basis, thus converting it into a matrix eigenvalue problem,
and by the Direct and Arnoldi methods integrated in the differential solvers
in Mathematica, see App.~\ref{sec:bogoliubov} for more details.

Both methods yield the same eigenvalues, and analyzing whether they
are real or complex, we split each region in stable and unstable parts.
\begin{figure}[t]
\centering
\includegraphics[width=.45\textwidth]{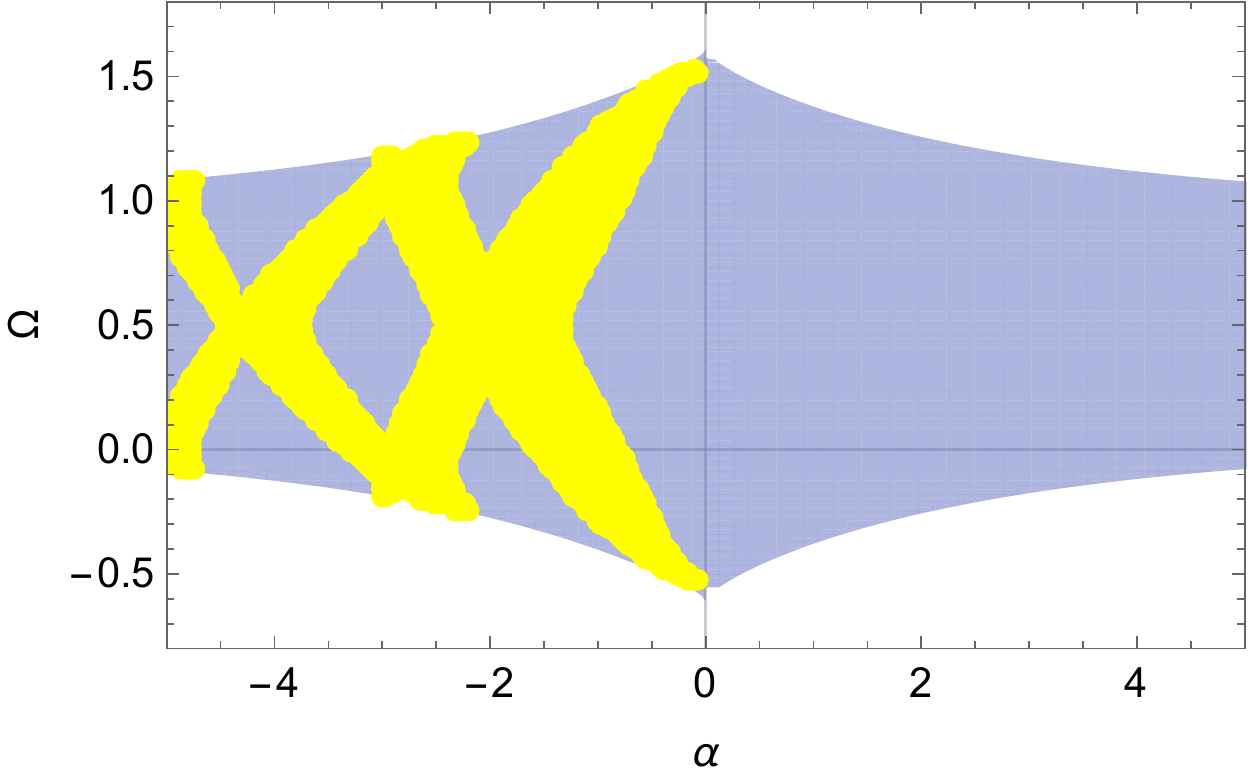}
\caption{(Color online). Metastable stripes appearing in the region corresponding
to the bottom of the second swallowtail diagram for $g=10$.}
\label{fig:metastability}
\end{figure}
The bottom of the first swallowtail and both upper parts  are found
completely stable and unstable, respectively.
In contrast, the lower part of
the second swallowtail is stable except for regions of metastability in form
of stripes in the plane $\alpha-\Omega$ at $\alpha<0$, as shown in Fig.~\ref{fig:metastability}.

\subsection{Dependence on nonlinearity}
\label{sec:g}

\begin{figure}[t]
\centering
\includegraphics[width=.48\textwidth]{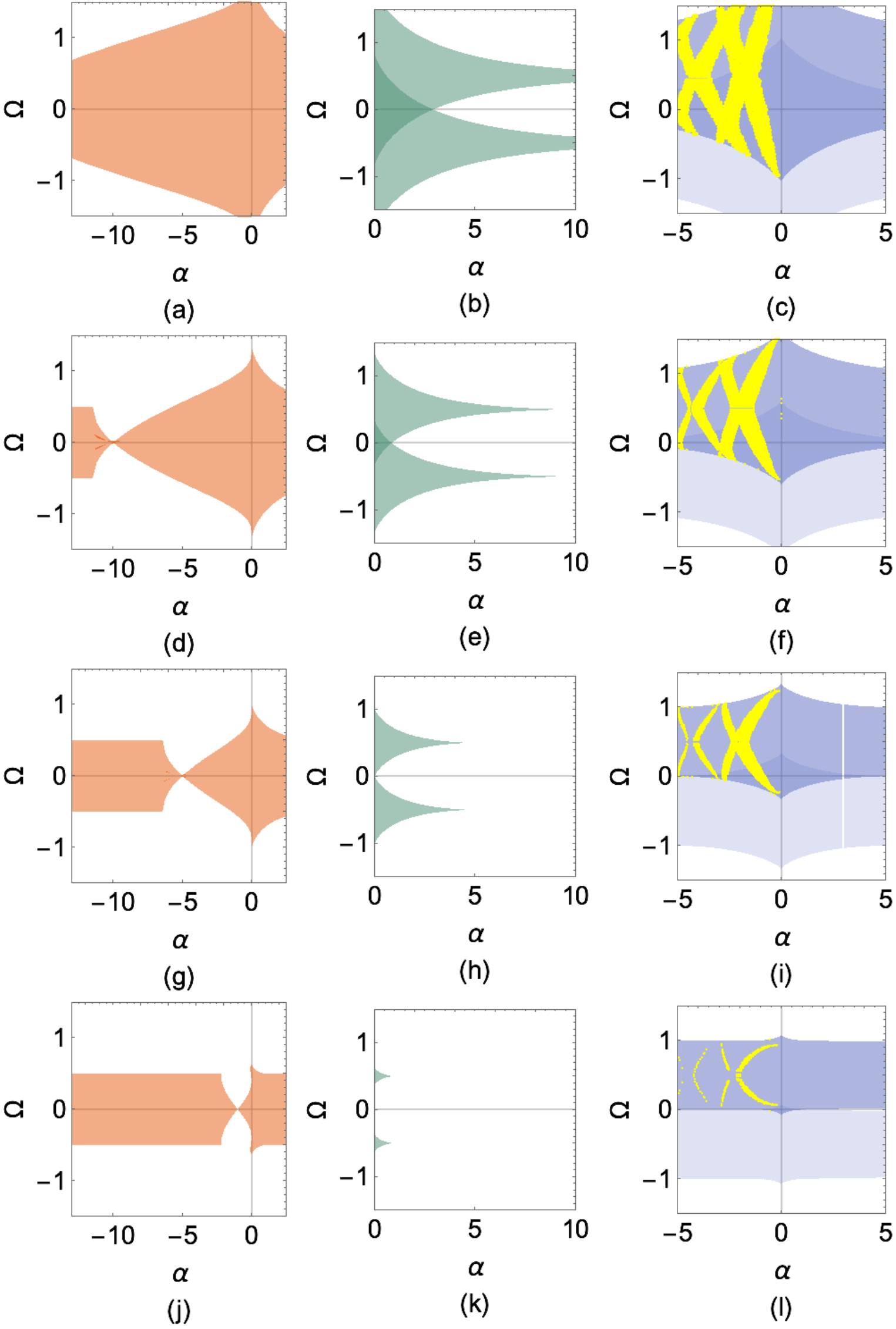}
\caption{Adiabatic regions corresponding to the bottom and
top of the first swallowtail diagram (first and second columns)
and bottom of the second one (third column) for $g=20,10,5$ and $1$
(from first to fourth row, respectively).
The plots in the third
column also contain the regions where solutions are found metastable, marked in yellow.
The second and third columns contain two of the possible regions
related through $\Omega\to\Omega+integer$. }
\label{fig:gdep}
\end{figure}

The results presented in the previous subsections have
been illustrated with the nonlinearity fixed to $g=10$.
In Fig.~\ref{fig:gdep} we show the same adiabatic regions
for $g=20,10,5$, and $1$. The first and third columns
correspond to the bottom part of first and second swallowtail diagrams (as in
Fig.~\ref{fig:sections}), while the middle
column represents the top part of the first one.
Larger nonlinearity implies a greater span and overlap of all the levels.
As $g$ decreases, and for any
fixed $\alpha$, the tail part
of the swallowtail diagrams becomes smaller, vanishing at $g=0$:
both the region in the middle column and the overlaps of shifted regions in the others decrease in size.
A condensate with larger nonlinearity is therefore able to sustain
stationary solutions for faster stirring velocities,
while in the linear limit, $g=0$, the critical velocities are independent
of the delta strength and fixed to $\Omega=\frac{l}{2}\pm\frac12$.

Performing the same metastability analysis we find that
the bottom part of the first swallowtail and both upper parts 
remain completely stable and unstable for all the $g$ tested.
The third region of Fig.~\ref{fig:gdep}, corresponding to the bottom of the second
swallowtail diagram, is also found completely stable at $\alpha>0$,
while the metastable stripes at $\alpha<0$ become thiner (thicker)
as $g$ decreases (increases).

\section{Adiabatic generation of vortex states and excited solitons}
\label{sec:excitation}

\begin{figure*}[t]
\centering
\includegraphics[width=.99\textwidth,height=240pt]{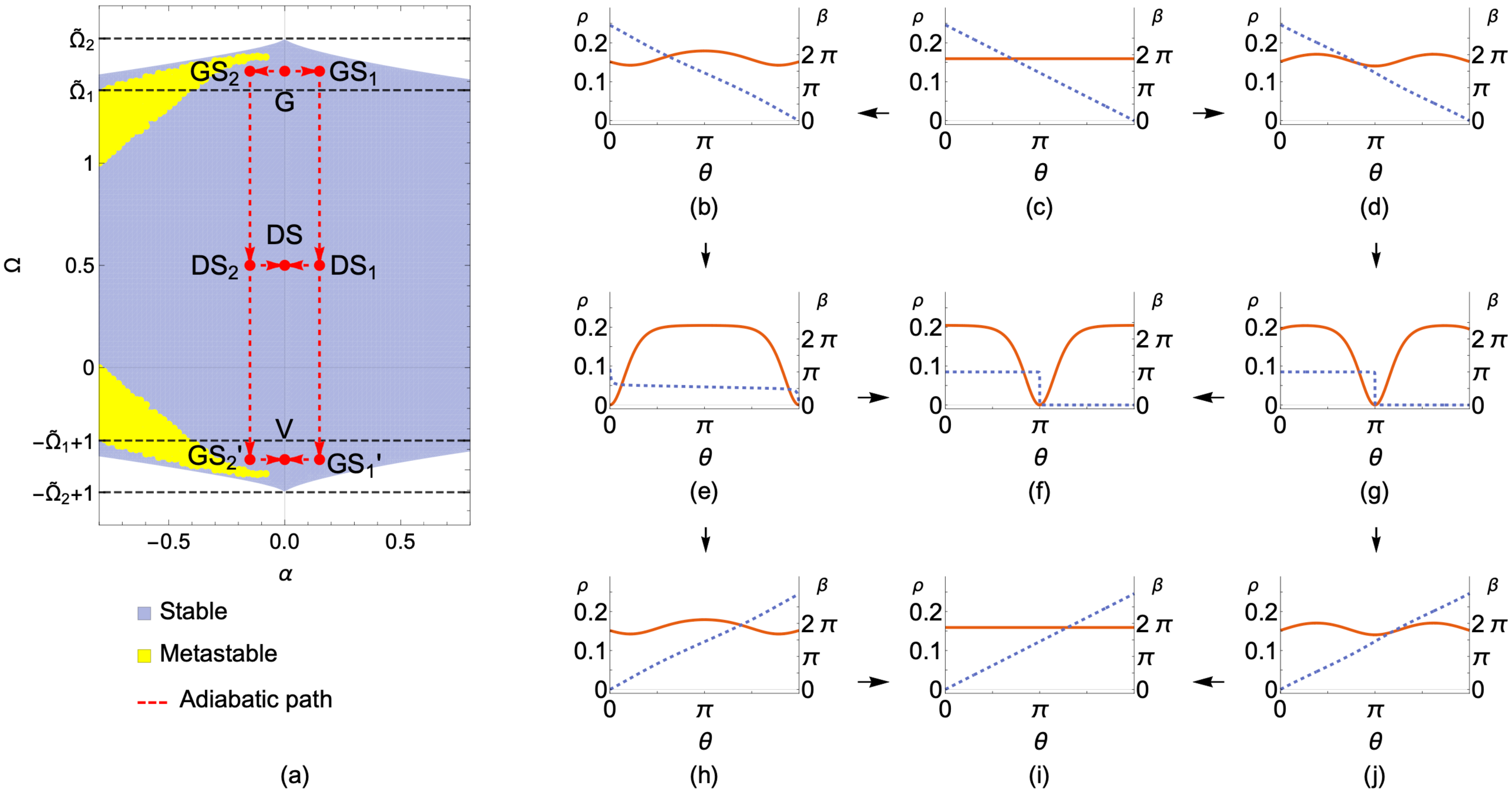}
\caption{(Color online). 
(a): four possible cycles that excite the condensate from the ground state (G)
to either a dark soliton (DS) or a vortex state with one quantum of angular momentum (V),
and with either a repulsive (subindex 1) or attractive (subindex 2) delta link.
(b)-(j): 
densities $\rho=|\phi|^2$ (red solid lines) and phases $\beta$ (blue dashed lines) in the comoving frame 
corresponding to  the vertices in the adiabatic paths
of plot (a).
All cycles start by setting a delta link while rotating at $\tilde{\Omega}_1<\Omega<\tilde{\Omega}_2$,
thus turning the ground state into a gray solitonic train with two depressions
((b) for $\alpha<0$ and (d) for $\alpha>0$).
Then the delta link is slowed down to $\Omega=\frac12$, where a dark soliton plus a kink is obtained
((e) and (g)).
As the delta is unset at this velocity, either repulsive or attractive, a dark soliton is obtained (f).
The density and phase profiles reached through an attractive delta are actually shifted $\Delta\theta=\pi$
with respect to plot (f).
If instead the velocity is further decreased to $-\tilde{\Omega}_2+1<\Omega<-\tilde{\Omega}_1+1$
((h) and (j)), and then
$\alpha$ is brought back to zero, the vortex state (i) is reached.
}
\label{fig:cycles}
\end{figure*}

All complex solutions found are continuously connected at
$\alpha=0$,
where gray solitonic trains merge into vortex states
as their velocity $\Omega$ departs from $\frac{l}{2}$, with $l$ an integer
(see middle panel of Fig.~\ref{fig:sections}).
Once a finite delta strength $\frac{\alpha}{2}\neq0$ is set,
a gap between the various swallowtail diagrams appears,
and depending on the initial rotational velocity and initial state,
different energy levels can be accessed.
In particular, setting a delta link in the ground state
while rotating at $\Omega_i\in(\tilde{\Omega}_n,\tilde{\Omega}_{n+1})$, $n\ge1$,
a gray solitonic train with $n+1$ depressions is obtained---reaching thus the bottom
of the $n+1$th swallowtail diagram.
As the velocity is decreased, the valleys become deeper,
and at $\Omega=\frac{n}{2}$ the solution turns into a wave train with $n$ dark solitons.
If the velocity is further decreased, the density profile becomes a
gray solitonic train again, with shallower waves with lower velocity.
At any velocity, one can unset the delta link, making the waves shallower as the delta strength
decreases back to $\alpha=0$.
For $\Omega>\tilde{\Omega}_n$, unsetting the link completely flattens the density,
and the ground state is recovered.
At velocities $\Omega\in(-\tilde{\Omega}_n+n,\tilde{\Omega}_n)$, the final state consists
in a gray solitonic train, with deeper waves the closer $\Omega$ is to $\frac{n}{2}$.
If one unsets the delta link at precisely $\Omega=\frac{n}{2}$,
a dark solitonic train is obtained.
In the case in which $\alpha$ is brought back to zero at $\Omega<-\tilde{\Omega}_n+n$,
the waves also become infinitely shallow, and merge with the vortex state
of $n$ quanta.

Cycles in which the final state is a vortex can be concatenated.
Once a vortex is produced, any observer can rotate at the same velocity as the condensate
current, such that in the comoving frame the vortex is observed as the ground state.
The cycle to produce a vortex from the ground state can then be repeated.
Let us consider the vortex state of five quanta of angular momenta.
It can be produced by setting a delta link rotating at
$\Omega_i\in(\tilde{\Omega}_5,\tilde{\Omega}_6)$, decreasing its velocity
to $\Omega_f\in(-\tilde{\Omega}_6+6,-\tilde{\Omega}_5+5)$, and then unsetting 
the delta. It involves a middle step in which a wave train with five dark solitons is produced.
Another possibility is to concatenate five analogous cycles in which
$\Omega_i\in(\tilde{\Omega}_1,\tilde{\Omega}_2)$, and
$\Omega_f\in(-\tilde{\Omega}_2+2,-\tilde{\Omega}_{1}+1)$, each involving the 
production of only one dark soliton. After each individual cycle,
the observer is boosted $\Omega\to\Omega+1$.

Any of these proposed paths should avoid the metastable and non-stationary
regions analyzed in the previous section.
In Fig.~\ref{fig:cycles}, we schematically demonstrate four of such adiabatic paths.
They excite the condensate
to one dark soliton and a vortex state, each one through the setting and unsetting of a delta link,
either repulsive or attractive.
In both cases, the delta strength is limited by the line
$\Omega_{cr}(\alpha)$ bounding the adiabatic region
corresponding to the bottom of the second swallowtail.
For the attractive delta, one also needs to avoid the metastable
region, which further limits the value of $\alpha$.
These adiabatic cycles can be reproduced for all $g$ tested. However, the range of
velocities at which the Dirac delta strength must be set,
decreases with $g$,
$\tilde{\Omega}_2-\tilde{\Omega}_1=\frac{3}{4}\sqrt{\frac{\pi}{2g}}+\mathcal{O}(g^{-\frac{3}{2}})$.
Moreover, in the case of the cycle with an attractive rotating
delta, the constrain on the magnitude of the delta potential
 will depend on the metastability stripes shown in Fig.~\ref{fig:gdep}.

For any adiabatic path involving higher swallowtail levels, the corresponding metastability analysis
and determination of critical velocity should be performed.
Alternatively, one can simulate these cycles by solving the time-dependent Gross-Pitaevskii
equation with the Dirac delta link replaced by a Gaussian one.
We solve this differential equation using the method of lines,
and reproduce the cycles proposed in Fig.~\ref{fig:cycles}. Moreover, we also
obtain dark solitonic trains and vortex states with up to five solitons or quanta.
On the one hand, these simulations validate the results found for the delta link.
On the other, they indicate that the structure of the spectrum laid out in the delta case
is also able to  depict the main features of the spectrum of a BEC stirred with a peaked Gaussian link.

The velocities
$\tilde{\Omega}_n=\sqrt{\frac{1}{2\pi}\frac{g}{g_{nat}}+\frac{n^2}{4}}\,\Omega_{nat}$,
constraining the adiabatic cycles,
have been presented in natural units, 
 where $\Omega_{nat}=\frac{\hbar}{MR^2}=1$ and
$\alpha_{nat}=g_{nat}=\frac{\hbar^2}{MR}=1$.
For a condensate of $^{87}$Rb atoms, with mass $M=86.909$ u 
and ring traps of radiuses $R_1=20{\,\rm\mu m}$ to $R_2=100{\,\rm\mu m}$,
the natural velocities range from
$\Omega_{nat}^{(1)}=1.83$ rad/s to $\Omega_{nat}^{(2)}=0.073$ rad/s.
Taking $g=10\,\tilde{g}$, the thresholds speeds to access the first excited states
are
$\tilde{\Omega}_n=2.48$, 2.95, 3.59 rad/s for the smaller radius, and
$\tilde{\Omega}_n=0.099$, 0.118, 0.143 rad/s for the larger one.
In the linear limit, $g=0$, $\tilde{\Omega}_n=\frac{n}{2}\Omega_{nat}$.

\section{Conclusions}
\label{sec:conclusions}

The spectrum of a 1D ring condensate with a Dirac delta rotating
at constant speed has been analyzed in terms of the nonlinearity $g$,
the delta velocity $\Omega$, and the delta strength $\alpha/2$.
Analytical expressions are provided for the wave function,
the current, and the chemical potential.
For a fixed $g$ and $\alpha$, the dependence of the chemical 
potential on the delta velocity, $\mu(\Omega)$,
consists of a series of swallowtail diagrams. These diagrams
can be organized from smaller to larger energies, and each one
can be split into a bottom and a top part. The lowest diagram
at $\alpha<0$ is an exception, and consists only of the bottom
part, with a more complex structure depending on a set
of critical points $P_i$. 
As the magnitude of the delta strength increases, the sizes of the tails
in each diagram decrease, while the energy gap among each diagram becomes larger.
At $\alpha=0$, the top parts of the diagrams merge with the
bottom parts of the immediate upper ones.
The spectrum $\mu(\alpha,\Omega)$ thus consists in a multiple
swallowtail 3D structure, each region providing a range
of delta strengths and velocities which can be varied
 adiabatically to access different solitonic solutions.
These solutions consist in gray or dark solitonic trains,
where the number of depressions in each train has been related
to the position of the swallowtail. In particular
an odd (even) number of dark solitons comove with the condensate
at $\Omega=\frac{l}{2}$, where $l$ is an odd (even) integer.

In order to support the possible adiabatic processes allowed
by the Gross-Pitaevskii spectrum, we have analyzed the metastability
of each solution for the first two swallowtail diagrams.
The top parts are found unstable and the bottom ones stable,
except for the bottom of the second diagram at $\alpha<0$,
which presents a series of metastable stripes in parameter space.

We have proposed a method to produce dark and gray solitons, and vortex states
of arbitrary quantized angular momentum,
by controlling the stirring velocity and strength of the potential.
The method consists in setting and unsetting a Dirac delta potential
while rotating it around the condensate at certain velocities.
In particular, as a rotating observer sets a delta link at
$\Omega_i\in(\tilde{\Omega}_n,\tilde{\Omega}_{n+1})$,
the bottom part of the $n+1$th swallow tail is reached,
and a gray solitonic train with $n+1$ depressions is produced.
The cycles in the parameter space defined by $\alpha$ and $\Omega$
corresponding to the various production processes
are constrained by the width of the swallowtail diagrams,
and also by the metastable regions.
These adiabatic paths are qualitatively reproduced by solving
the time-dependent Gross-Pitaevskii equation in which a finite width Gaussian
link is rotating at constant speed.

\appendix
\label{app:model}
\section{Gross-Pitaevskii equation and boundary conditions}
\label{sec:gpe}
The evolution of the condensate wave function
in the Lab frame, $\psi_L(\theta_L,t_L)$, is governed by the 1D
Gross-Pitaevskii equation,
\begin{align}
i\hbar\,\partial_{t_L}\psi_L=&
-\frac{\hbar^2}{2M R^2}\,\partial_{\theta_L}^2\psi_L
+g|\psi_L|^2\psi_L
\nonumber\\&+\frac{\alpha}{2}\,\delta(\theta_L-\Omega\,t_L)\psi_L,
\label{eq:gp1}
\end{align}
where $M$ is the atomic mass, $R$ the radius of the ring,
$\theta_L\in(0,2\pi)$ and $t_L$ the angular and time coordinates in the lab frame,
$\alpha/2$ and $\Omega$ the magnitude and angular velocity of the Dirac delta,
and $g>0$ the reduced 1D coupling strength.
The circular topology imposes continuity conditions in the wave function,
\begin{align}
\label{eq:contc}
\psi_L(\Omega t_L,t_L)=&\psi_L(\Omega t_L+2\pi,t_L),
\end{align}
and the Dirac delta constrains its derivatives through boundary conditions.
These are obtained by integrating Eq.~(\ref{eq:gp1}) in a small contour around the delta, $\theta_L\in(\Omega t_L-\epsilon,\Omega t_L+\epsilon)$, and taking the limit $\epsilon\to0$,
\begin{align}
\label{eq:deriv}
\frac{m R^2\alpha}{\hbar^2}\,\psi(\Omega t_L,t_L)
=&
(\partial_{\theta_L}\psi)|_{\theta_L=\Omega t_L}
\\\nonumber
&-(\partial_{\theta_L}\psi)|_{\theta_L=\Omega t_L+2\pi}.
\end{align}
We change variables to the delta rotating frame \cite{leggett73},
\begin{align}
&\theta=\theta_L-\Omega\,t_L,~~~~~\partial_{\theta}=\partial_{\theta_L},
\\
&t=t_L,~~~~~~~~~~~~~~~\partial_{t}=\partial_{t_L}+\Omega\,\partial_{\theta_L},
\\
&\psi(\theta,t)=e^{\frac{i}{\hbar} \left(\frac12mR^2\Omega^2t_L-m R\,\Omega \theta_L\right)}\psi_L(\theta_L,t_L),&
\end{align}
and use units $\hbar=M=R=1$.
Then eqs.~(\ref{eq:gp1}), (\ref{eq:contc}), and (\ref{eq:deriv}) become,
\begin{align}
i\,\partial_t\psi=&
-\frac{1}{2}\,\partial_\theta^2\psi+g|\psi|^2\psi,
\label{eq:gp2}\\
\psi(0,t)=&
e^{i \Omega \theta}\psi(2\pi,t),
\\
\psi(0,t)
=&
\frac{1}{\alpha}\Big[
(\partial_\theta\psi)|_{\theta=0}
-e^{i \Omega \theta}
(\partial_\theta\psi)|_{\theta=2\pi}\Big].
\end{align}
For a stationary solution, $\psi(\theta,t)=e^{-i\,\mu t}\phi(\theta)$,
where $\mu$ is the chemical potential, these equations result in
Eqs.~(\ref{eq:gpf})-(\ref{eq:bc2}).

\section{Solutions}
\label{sec:solutions}
\begin{figure}[t]
\centering
\includegraphics[width=.45\textwidth,height=100 pt]{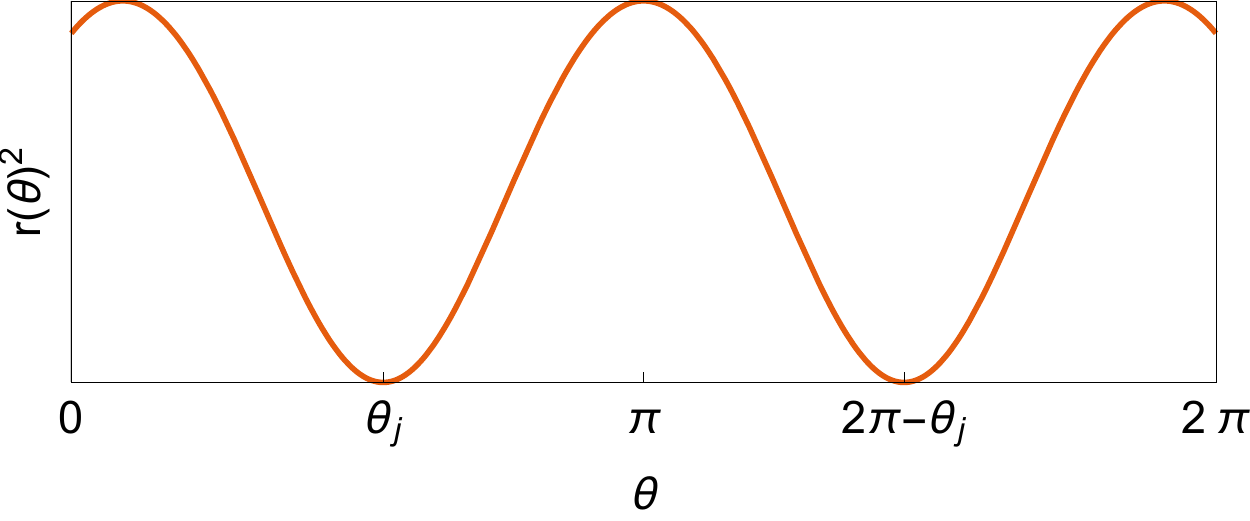}
\caption{Example of a density $r(\theta)^2$ with a shift $\theta_j$ such that satisfies
periodic boundary conditions at $\theta=0,2\pi$.}
\label{fig:shift}
\end{figure}

To obtain the solutions of Eq.~(\ref{eq:gpf}) we
write the wave function as
$\phi(\theta)=r(\theta)e^{i\,\beta(\theta)}$.
Separating into real and imaginary parts, and integrating,
the density and phase take the general form,
\begin{align}
\label{eq:density}
r^2_J(\theta)=&A+B\,J^2(k(\theta-\theta_j),m),
\\
\label{eq:phase}
\beta'_J(\theta)=&\frac{\gamma_J}{r_J^2(\theta)},
\end{align}
where $J$ is one of the 12 Jacobi functions and
$A$, $B$, $k$, $\theta_j$, $m$, and $\gamma_J$ are constants.
The squares of the six convergent or divergent Jacobi functions
are related among themselves linearly and through shifts in $\theta$,
and therefore one may consider only a convergent one and a divergent one
with general $A$, $B$, and $\theta_j$.
Eq.~(\ref{eq:phase}) represents the stationarity condition,
with $\gamma_J=r_J^2\beta_J'$ the current.
The shift $\theta_j$ is fixed by the continuity condition, $r(0)=r(2\pi)$:
the angular length of the condensate, $2\pi$, has to be equal to 
an integer number of periods ($j\,T$) plus twice the shift,
$2\pi=j\,T+2\theta_j$ (see Fig.~\ref{fig:shift}).
The period of $r(\theta)$ is given in terms of the elliptic integral of first kind
($K(m)$), $T=\frac{2K(m)}{k}$, and therefore
\begin{align}
\theta_j=\pi-\frac{j}{k}K(m).
\end{align}
Since we take $k$ and $m$ as a parameters, $j$ can be fixed to $j=0,1$.
Eqs.~(\ref{eq:gpf}) and~(\ref{eq:norm}) fix
$A$, $B$, $\gamma_J$ and $\mu_J$ in terms of $\theta_j$, $k$, and $m$.
Using the Jacobi functions dn and dc as the convergent
and divergent independent solutions, respectively,
the amplitudes read,
\begin{align}
\label{eq:amp}
r_{{\rm dn}}(\theta)=&\frac{\sqrt{g+k\,\eta_{{\rm dn}}-2\pi k^2{\rm dn}^2(k(\theta-\theta_j),m)}}{\sqrt{2\pi g}},
\\
r_{{\rm dc}}(\theta)=&\frac{\sqrt{g+k\,\eta_{{\rm dc}}-2\pi k^2-2\pi k^2{\rm dc}^2(k(\theta-\theta_j),m)}}{\sqrt{2\pi g}},
\end{align}
where
\begin{align}
\eta_{{\rm dn}}=&
E[{\rm JA}(k(2\pi-\theta_j),m),m]
\nonumber\\&
+E[{\rm JA}(k\,\theta_j,m),m],
\\
\eta_{{\rm dc}}=&\eta_{{\rm dn}}+{\rm dn}(k\theta_j){\rm sc}(k\theta_j),
\end{align}
with $E$ the elliptic integral of the second kind,
${\rm JA}$ the Jacobi amplitude, sc the Jacobi function, and where dc allows 
only for $j=0$ and $k<K(m)/\pi$ in order to be convergent
in $\theta\in[0,2\pi)$.
The phases $\beta_{{\rm dn}}$ and $\beta_{{\rm dc}}$ are then integrated from 
the Eq.~(\ref{eq:phase}).

For the types of Jacobi functions chosen, $J={\rm dn}, {\rm dc}$,
the corresponding currents and chemical potentials read,
\begin{align}
\gamma_{J}=&\frac{\pm1}{g(2\pi)^{3/2}}\sqrt{g+k\,\eta_J}\sqrt{g-2\pi k^2+k\,\eta_J}
\nonumber\\&\times\sqrt{g-2\pi k^2(1-m)+k\,\eta_J},
\label{eq:gamma}
\\
\label{eq:mu}
\mu_{J}=&\frac{1}{4\pi}\left(3g+2k^2(m-2)+3k\,\eta_J\right).
\end{align}
This leaves the frequency $k$ and elliptic modulus $m$
as the only free parameters. They constrain $\alpha$ and $\Omega$ through
the boundary conditions in Eqs.~(\ref{eq:bc1}) and~(\ref{eq:bc2}).
$k$ and $m$ are either real, $k>0$, $m\in[0,1]$, or,
in the case of real solutions, may also take complex values with $|m|=1$
and $k\propto 1/\sqrt{1+m}$. For the real solutions
(with general real boundary conditions) we refer to~\cite{perezobiol19}.
The elliptic modulus is further constrained by the condition that $\gamma_J\in\mathcal{R}$,
which is satisfied when and odd number of radicants in Eq.~\ref{eq:gamma} are positive.
Note that the transition to an even number of radicants being negative, where $\gamma_J$
is not real, happens at $\gamma_J=0$.

In the case of $\alpha=0$, the wave functions are plane waves,
where $\mu_{n}=\frac{g}{2\pi}+\frac12(\Omega+n)^2$,
or solitonic trains,
which have, for each level $n$,
\begin{align}
\mu_{\alpha=0}=&\frac{1}{4\pi}\left[
3g+\frac{2(n-1)^2(m-2)}{\pi}K(m)^2
\nonumber
\right.\\&\left.+\frac{6(n-1)^2}{\pi}K(m)E(m)
\right].
\end{align}
Both expressions correspond to periodic boundary conditions
(solved in Ref.~\cite{carr002}) and coincide at $m=0$ and $k=\frac{n-1}{\pi}K(0)=\frac{n-1}{2}$.
These values determine the critical velocity,
\begin{align}
\label{eq:omcr0}
\tilde{\Omega}_n=\sqrt{\frac{g}{2\pi}+\frac{n^2}{4}},
\end{align}
that bounds both regions $n$ and $n+1$ at $\alpha=0$.

For a similar treatment of the Jacobi functions
 and a complete derivation of the solutions see e.g.~\cite{carr002}.
The main difference between \cite{carr002} and our work is that in \cite{carr002} $k$ and $m$ 
are not taken as parameters to account for a phase jump and a kink,
but adjusted such that periodic boundary conditions are obtained.

\begin{figure}[t]
\centering
\includegraphics[width=.45\textwidth,height=248pt]{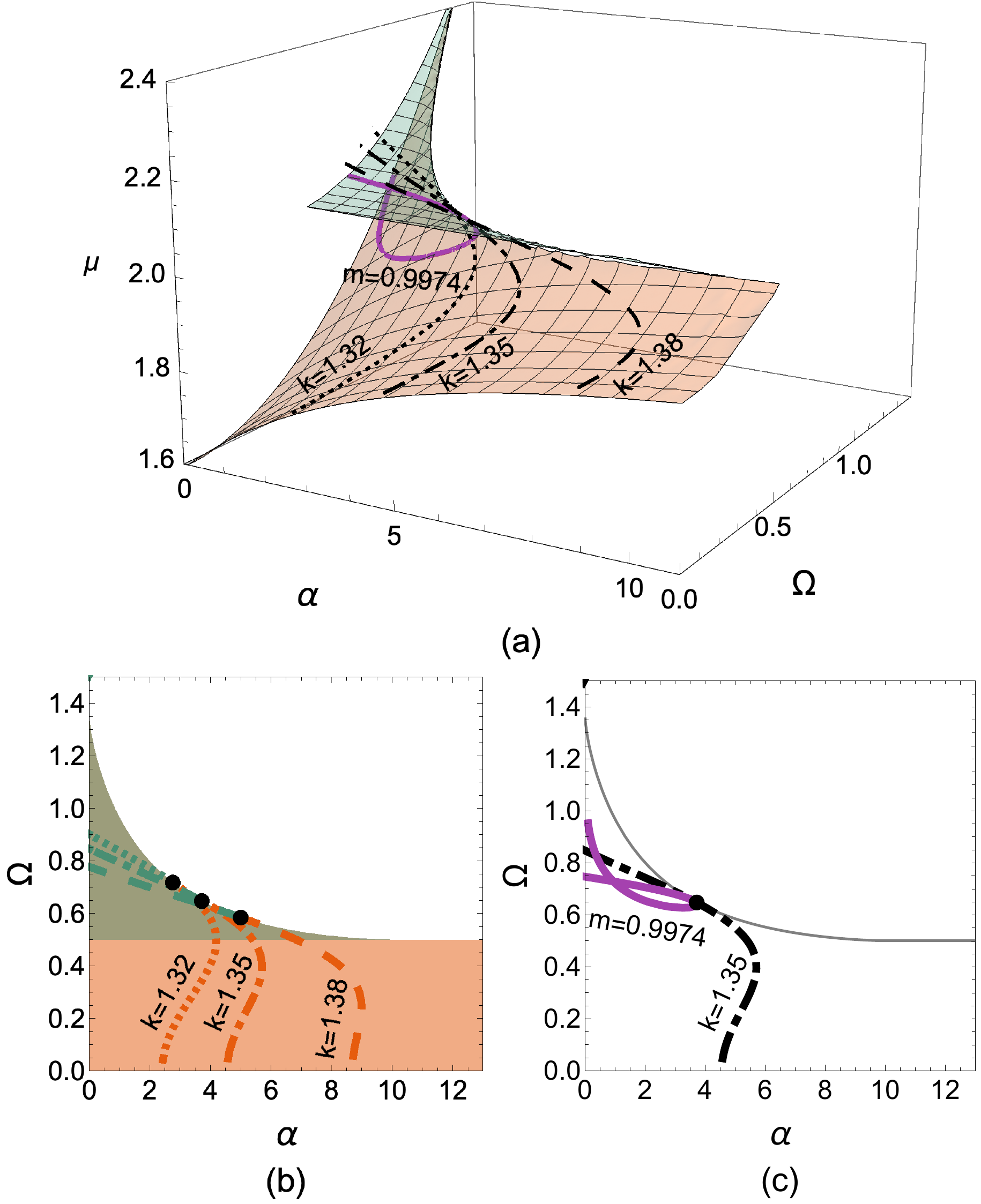}
\caption{(Color online)
(a): lower part of the spectrum $\mu(\alpha,\Omega)$ for $g=10$ obtained
by scanning the frequency $k>0$ and the elliptic modulus $m$ through its parametric definition,
$(\alpha(k,m),\Omega(k,m),\mu(k,m))$.
Three dashed lines in which the frequency has been fixed to $k=1.32,1.35$ and $1.38$
and parametrized by $m$ have been drawn in order to show how the spectrum
can be computed in a systematic way. 
Alternatively, one can first fix $m$ and then run $k$, obtaining lines as the purple solid curve
(for $m=0.9974$).
The ground state (red bottom surface) and the first excited state (green top surface),
become degenerate at a line in which the curves parametrized by $k$ (solid purple)
and $m$ (black dashed) become tangent.
To obtain the complete spectrum these surfaces are translated and mirrored at $\Omega\to\pm\Omega+integer$.
(b): projection of figure (a) in the $\alpha-\Omega$ plane,
where the three lines parametrized by $m$ are also included. The part of these lines
spanning the ground state is colored red, while the one corresponding to the 
top level is colored green.
(c): the degeneracy or critical line (light gray) can be computed
by constraining the curves parametrized with  $k$ and $m$ to be tangent to each other,
as the purple solid and dashed black ones plotted in the figure.
}
\label{fig:mapping}
\end{figure}

\section{Computation of the spectrum}
\label{sec:computation}

All solutions $\phi=r\,e^{i\,\beta}$ 
satisfying Eqs.~(\ref{eq:gpf})-(\ref{eq:norm}) can be obtained by running
$k$ and $m$ in their allowed ranges in any of the three Jacobi functions,
two convergent and one divergent, and which we label as
\begin{align}
\phidnz=&r_{{\rm dn}}^{(j=0)}(\theta)e^{i\beta_{{\rm dn}}^{(j=0)}},
\\ 
\phidno=&r_{{\rm dn}}^{(j=1)}e^{i\beta_{{\rm dn}}^{(j=1)}},
\\
\phidc=&r_{{\rm dc}}^{(j=0)}e^{i\beta_{{\rm dn}}^{(j=0)}}.
\end{align}
Then, for each $k$ and $m$, the delta strength, velocity, and the chemical potential are obtained
from Eqs.~(\ref{eq:bc1}),~(\ref{eq:bc2}), and~(\ref{eq:gpf}),
\begin{align}
\label{eq:alpha}
\alpha_J(k,m)=&\frac{r_J'(0)-r_J'(2\pi)}{r_J(0)},
\\\label{eq:omega}
\Omega_J(k,m)=&\frac{1}{2\pi}[\,\beta_J(2\pi)-\beta_J(0)\,],
\\
\mu(k,m)=&\frac{1}{\,\phi(0)}
\left(-\frac12 \phi''(0)+g|\phi(0)|^2\phi(0)\right).
\label{eq:mu2}
\end{align}
A sample of the spectrum $\mu(\alpha,\Omega)$ produced this way
is shown in plot (a) of Fig.~\ref{fig:mapping}. Its projection on
the $\alpha-\Omega$ plane is potted on panel (b) of the same figure.
In this plane, lines parametrized by $k$ and fixed $m$,
($\alpha_m(k)$,$\Omega_m(k)$), and lines parametrized by $m$ and fixed $k$,
($\alpha_k(m)$,$\Omega_k(m)$), are tangent at the degeneracy line,
\begin{align}
\pard{k}\left(\alpha_m(k),\Omega_m(k)\right)
\propto
\pard{m}\left(\alpha_k(m),\Omega_k(m)\right),
\end{align}
see bottom-right panel of Fig.~\ref{fig:mapping}.
Therefore, any degeneracy line may be computed by solving 
\begin{align}
\pard[\Omega(k,m)]{k}\pard[\alpha(k,m)]{m}
=
\pard[\Omega(k,m)]{m}\pard[\alpha(k,m)]{k}.
\label{eq:criteq}
\end{align}

\begin{figure*}[t]
\centering
\includegraphics[width=.97\textwidth,height=140pt]{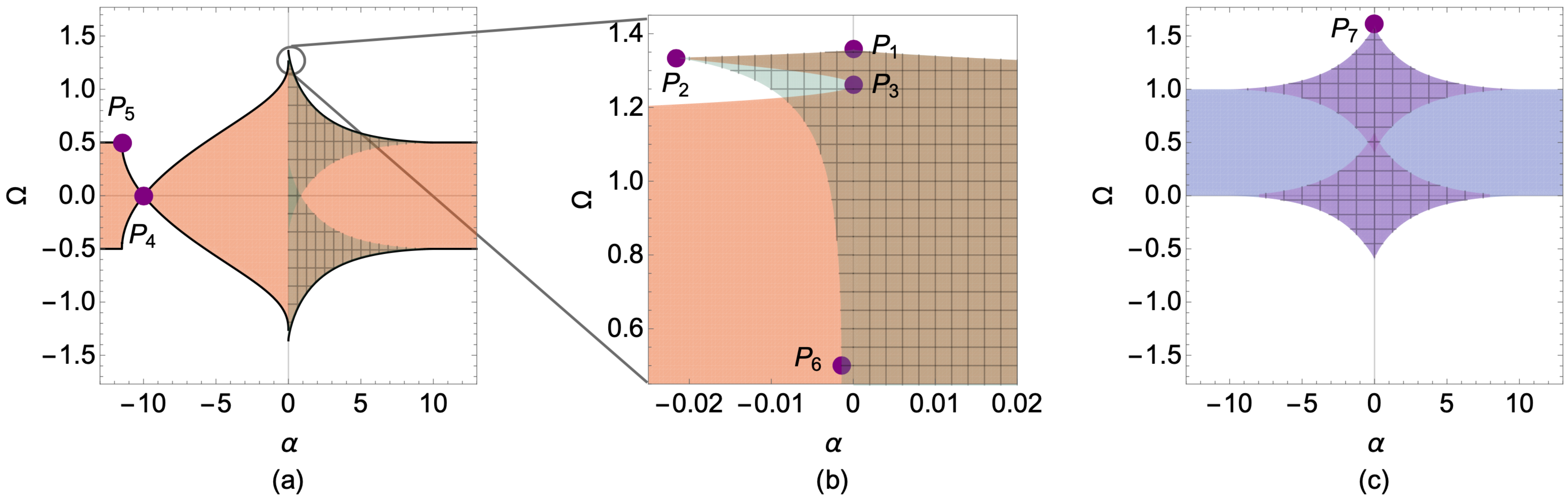}
\caption{(Color online).
Regions in which solutions are adiabatically connected through a variation of the delta 
strength $\alpha/2$ and velocity $\Omega$ for $g=10$.
They correspond to the bottom (red solid) and top (green grid) of the first swallowtail
diagram, (plots (a) and (b)), and the bottom (blue solid) and top (purple grid) 
of the second one (plot (c)).
Points $P_i$, $i=1,\cdots,7$ characterize the structure of the first energy levels at $\alpha<0$.
Due to rotational symmetry, any of these regions can be shifted $\Omega\to\Omega+integer$,
as in both surfaces with grids.}
\label{fig:regions}
\end{figure*}

\section{Spectral structure}
\label{sec:structure}

Fig.~\ref{fig:regions} shows the projections of $\mu(\alpha,\Omega)$ in the $\alpha-\Omega$ plane
for the ground state and first excited levels.
The ground and first excited state for $\alpha<0$ present a more complex structure, which we analyze in this Appendix.
Its bounds are determined by the limits of the elliptic modulus $m$
---0, 1, or the ones fixed by $\gamma_J=0$---, or by Eq.~(\ref{eq:criteq}),
and conform the set of lines uniting the points
$P_1-(\cdots)-P_5$ and $P_1-P_2-P_6$.
Each segment of these bounds is listed
in Table~\ref{tab:bounds} together with the function and limit of $m$ they represent.
The line $P_2-(\cdots)-P_5$ and its continuation at $\Omega=\frac12$
is the only curve not defining the degeneracy of two energy levels.
The points $P_i$, $i=1,7$, themselves can be further determined.
Points $P_1$, $P_3$, and $P_7$ correspond to $\alpha=0$ and velocities
$\tilde{\Omega}_1$, $\tilde{\Omega}_0$, and $\tilde{\Omega}_2$, respectively.
The values of $P_4$, $P_5$ are constrained by their current
and elliptic modulus being zero, $\gamma_J=m=0$, while
$P_6$ has $\gamma_J=0$ and $m\to1$. Point $P_2$ is obtained
by minimizing $\alpha$ with $m=1$. All these constrains
fix $P_i$ to the values shown in Table \ref{tab:points}.

\begin{table}[t]
\renewcommand{\arraystretch}{1.2}
\begin{tabular}{c|c|c|c}
\multicolumn{2}{c|}{Region}                                                                         & $J$                   & Bounds \\\hline
\multirow{4}{40pt}{Bottom 1\textsuperscript{st} ST $\alpha<0$}  & \multirow{2}{*}{$P_1-P_2-P_3$}    & \multirow{2}{*}{dn}   & $P_1-P_2$, Eq.~(\ref{eq:criteq}) \\\cline{4-4}
							                                  	& 								    &						  & $P_2-P_3$, $m=1$    \\\cline{2-4}
                                                                & \multirow{2}{40pt}{$P_3-P_4-P_5-P_\infty$}    & \multirow{2}{*}{dc}   & $P_3-P_4-P_5$, $m=0$ \\\cline{4-4}
                                                                &									&						  & $P_5-P_\infty$, $\gamma=0$ \\\hline
\multicolumn{2}{c|}{Bottom 1\textsuperscript{st} ST $\alpha>0$}                                     & $\rm\tilde{dn}$      & Eq.~(\ref{eq:criteq}) \\\hline
\multicolumn{2}{c|}{\multirow{2}{*}{Top 1\textsuperscript{st} ST $\alpha<0$}}                       & \multirow{2}{*}{dn}   & $P_1-P_2$, Eq.~(\ref{eq:criteq}) \\\cline{4-4}
\multicolumn{2}{c|}{}		                                  										&						  & $P_2-P_6$, $m=1$    \\\hline
\multicolumn{2}{c|}{Top 1\textsuperscript{st} ST $\alpha>0$}                                        & $\rm\tilde{dn}$       & Eq.~(\ref{eq:criteq}) \\\hline
\multicolumn{2}{c|}{Bottom 2\textsuperscript{nd} ST $\alpha<0$}                                     & $\rm\tilde{dn}$       & Eq.~(\ref{eq:criteq}) \\\hline
\multicolumn{2}{c|}{Bottom 2\textsuperscript{nd} ST $\alpha>0$}                                     &  dn                   & Eq.~(\ref{eq:criteq}) 

\end{tabular}
\caption{Relation between the adiabatic regions, the Jacobi function with which the solutions in the
region are computed, and the constraints of the function at the boundaries.
The regions are defined according to the swallowtail (ST) structure of Fig.~\ref{fig:sections} and
through the lines bounding them. In the first swallowtail at $\alpha<0$, these lines are described by the union of the various points $P_i$ as in Fig.~\ref{fig:regions}, and where $P_\infty\equiv(\alpha\to-\infty,\Omega=\frac12)$.}
\label{tab:bounds}
\end{table}

\begin{table}[t]
\begin{tabular}{c|c|c}
  $P_i$    & $\alpha_{P_i}$    & $\Omega_{P_i}$  \\\hline
$P_1$  & $0$  & $\sqrt{\frac{g}{2\pi}+\frac{1}{4}}$ \\\hline
$P_2$  & $\begin{aligned}&~~~~~~~\alpha_{{\rm dn}}(k_{2},1),\\
&2 k_2 \left(-3\pi g+8\pi^2 k_2^2+2\right)\\
&+\left(4 \pi  k_2^2-3 g\right) \sinh (4 \pi  k_2)\\
&+2 (\pi  g-2) k_2 \cosh (4 \pi  k_2)\\
&-6 g \sinh (2 \pi  k_2)\\
&-4 \pi  g k_2 \cosh (2 \pi  k_2)=0
\end{aligned}$  & $\Omega_{P_2}=\Omega(k_{2},1)$ \\\hline
$P_3$  & $0$  & $\sqrt{\frac{g}{2\pi}}$ \\\hline
$P_4$  & $-g$  & $0$ \\\hline
$P_5$  & $\begin{aligned}&~~~~\alpha_{{\rm dc}}(k_{5},1),\\ &-2k_{5}\tan(k_{5}\pi)\\&+g+2\pi k_{5}^2=0\end{aligned}$  & $\frac12$ \\\hline
$P_6$  & $\begin{aligned}&~~~~\alpha_{{\rm dn}}(k_{6},1),\\&2k_{6}\tanh(k_{6}\pi)\\&+g-2\pi k_{6}^2=0\\\end{aligned}$  & $\frac12$  \\\hline
$P_7$  & $0$  & $\sqrt{\frac{g}{2\pi}+1}$ \\
\end{tabular}
\caption{Expressions of $\alpha$ and $\Omega$ for the critical
points $P_i=(\alpha_i,\Omega_i)$, $i=1,\cdots,7$.}
\label{tab:points}
\end{table}

\begin{figure}[t]
\centering
\includegraphics[width=.47\textwidth]{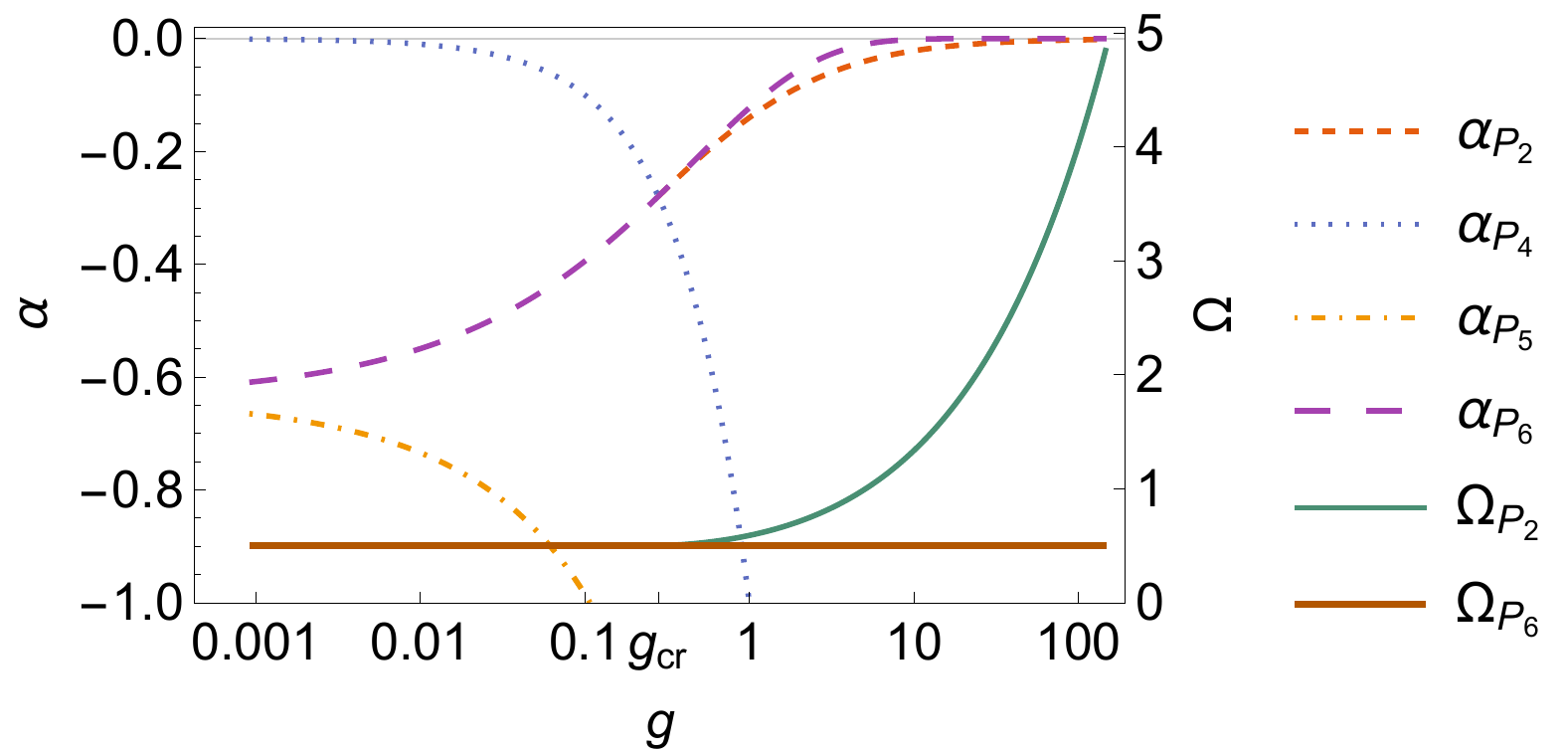}
\caption{(Color online). Dependence of the critical points $P_i=(\alpha_{P_i},\Omega_{P_i})$
 on the nonlinearity $g$. The plots are only for the values of $\alpha_{P_i}(g)$ and $\Omega_{P_i}(g)$
which do not have closed analytical expressions, and also for $\alpha_{P_4}$ and
$\Omega_{P_6}$ for comparison.
}
\label{fig:pts}
\end{figure}

The dependence of the spectrum on $g$ can be analyzed quantitatively
through the expressions in Table~\ref{tab:points} for the
points $P_i$. $P_1$, $P_3$, $P_4$, and $P_7$ are given
in analytical form, and $P_2$, $P_5$, and $P_6$ are
plotted in Fig.~\ref{fig:pts}. $\alpha_{P_2}$ approaches zero as $g$
increases, and the structures bounded by the lines $P_1-P_2-P_3$
and $P_1-P_2-P_6$ (middle panel in Fig.~\ref{fig:regions}) vanish
in the limit $g\to\infty$.
In contrast, $|\alpha_{P_4}|$, $|\alpha_{P_5}|$ and $\Omega_{P_3}$ increase with 
$g$, and the region bounded by these points grows at large interactions.
At $g=-\frac{2}{\pi}+2\pi k_{cr}^2\simeq 0.280$,
where $k_{cr}\simeq 0.382$ solves $\pi k_{cr}\tanh(k_{cr}\pi)=1$,
both Eqs. for $k_2$ and $k_6$ in Table~\ref{tab:points} are satisfied
and points $P_2$ and $P_6$ coincide.
In the limit $g\to0$, $P_6$ approaches $P_5$ at
$\alpha=-\frac{2}{\pi}$, $k_{5}$ and $k_{6}$ tend to zero,
and $\Omega_{P_1}=\frac12$: the parts of the region merge into a flat band. In general, at $g=0$, all levels
turn into regions spanning $\Omega\in[l,l+1]$,
where all the solutions are stable (see App.~\ref{sec:linear} for
the linear solutions).

\section{Bogoliubov analysis}
\label{sec:bogoliubov}

The differential equations and boundary conditions constraining $\tilde{u}$ and $\tilde{v}$ are
\begin{align}
w\,\tilde{u}=&-\frac12\tilde{u}''+i\,\Omega\tilde{u}'+\frac12\,\Omega^2\tilde{u}
\\&\nonumber
+2g\,|\phi|^2\tilde{u}-\mu\,\tilde{u}-g\phi^2\tilde{v},
\\
w\,\tilde{v}=&\frac12\tilde{v}''-i\,\Omega\tilde{v}'-\frac12\,\Omega^2\tilde{u}
\\&\nonumber
-2g\,|\phi|^2\tilde{v}+\mu\,\tilde{v}+g{\phi^*}^2\tilde{u},
\end{align}
\begin{align}
\label{eq:bcu1}
\tilde{u}(0)-\tilde{u}(2\pi)=&0,
\\
\label{eq:bcu2}
\tilde{u}'(0)-\tilde{u}'(2\pi)=&\alpha\,\tilde{u}(0),
\\
\label{eq:bcv1}
\tilde{v}(0)-\tilde{v}(2\pi)=&0,
\\
\label{eq:bcv2}
\tilde{v}'(0)-\tilde{v}'(2\pi)=&\alpha\,\tilde{v}(0).
\end{align}
In order to turn this system of equations into a linear eigenvalue problem,
we expand $\tilde{u}$ and $\tilde{v}$ in an orthonormal basis.
This basis does not consist in periodic plane waves, since the derivatives
must be discontinuous according to the delta conditions,
but in the solutions of Eqs.~(\ref{eq:gpf})-(\ref{eq:norm}) with $g=0$ and $\Omega=0$.
Imposing these constraints
on exponential and trigonometric functions, we obtain the basis,
\begin{align}
s_0(\theta)=&\frac{e^{k_0(2\pi-\theta)}+e^{k_0\theta}}{\sqrt{k_0/(-1+e^{4\pi k_0}+4\pi k_0e^{2\pi k_0})}},
\\
s_{2n+1}(\theta)=&\frac{\cos(k_n(\theta-\pi))}{\sqrt{\pi+\sin(2\pi k_n)/(2k_n)}},
\\
s_{2n}(\theta)=&\frac{\sin(n\,\theta)}{\sqrt{\pi}},
\end{align}
with $n$ a positive integer, and where the element $s_0(\theta)$ is only used for $\alpha<0$.
$\tilde{u}$ and $\tilde{v}$ expanded in this set of functions solve
Eqs.~(\ref{eq:bcu1}) and~(\ref{eq:bcv1}),
and Eqs.~(\ref{eq:bcu2}) and~(\ref{eq:bcv2})
are satisfied as long as $k_0$ and $k_n$ are the solutions of,
respectively,
\begin{align}
\alpha=&2\,k_0\,\frac{e^{2\pi k_0}-1}{e^{2\pi k_0}+1},
\\
\alpha=& 2\,k_n \tan(k_n\,\pi).
\end{align}

\section{Linear limit}
\label{sec:linear}

\begin{figure}[t]
\centering
\includegraphics[width=.49\textwidth,height=90pt]{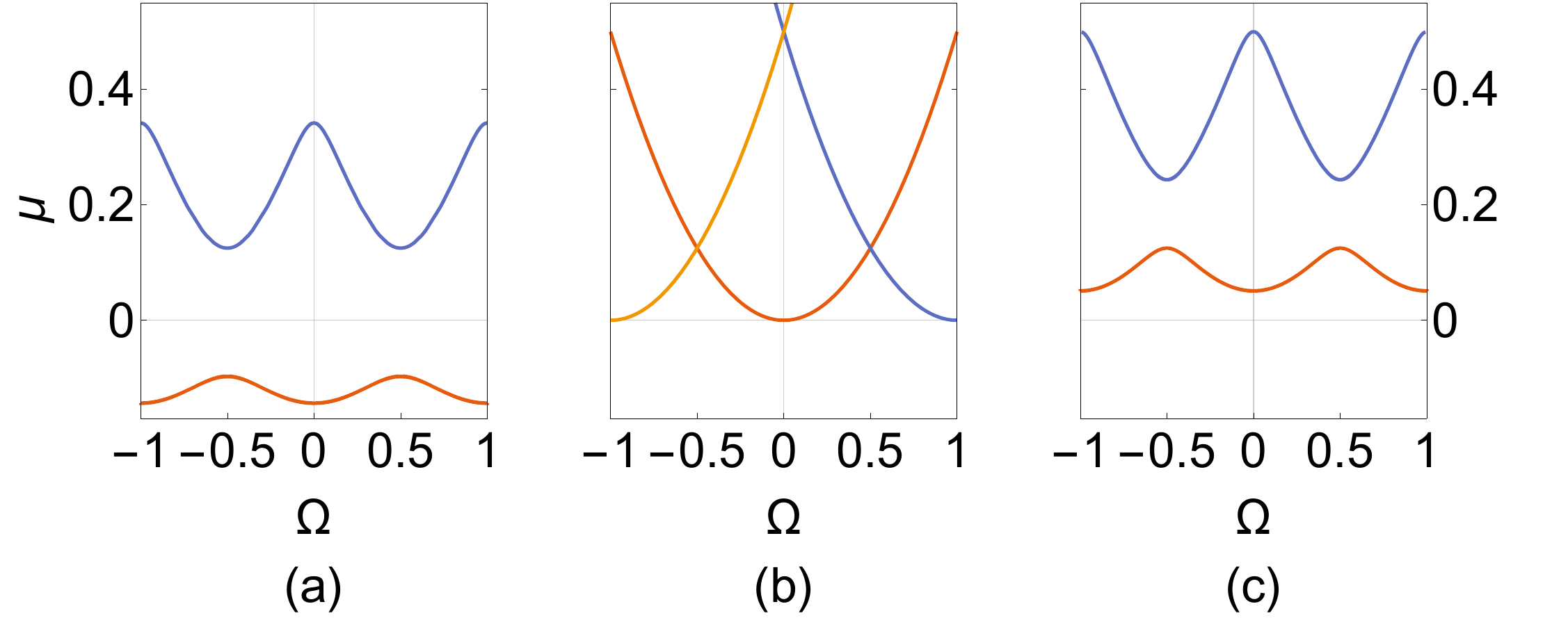}
\caption{(Color online). Sections $\alpha=-1,0$ and $1$ ((a), (b), and (c)) of the first two energy
levels in the spectrum $\mu(\alpha,\Omega)$ for $g=0$.}
\label{fig:g0}
\end{figure}

Solutions of Eqs.~(\ref{eq:gpf})-(\ref{eq:norm}) with $g=0$ can be found analytically
proceeding analogously to App.~\ref{sec:solutions} and replacing
Jacobi functions by trigonometric and hyperbolic ones. They read,
\begin{align}
r_{c}^2=&A_{c}\left[1+B_{c}\,\cos(k(\theta-\pi))^2\right],
\\
r_{ch}^2=&A_{ch}\left[1+B_{ch}\,\cosh(k(\theta-\pi))^2\right],
\end{align}
where,
\begin{align}
 A_{c}=&\frac{\gamma^2\left[2\pi k+\sin(2\pi k)\right]}
{k^3\pm k\sqrt{k^4-(2\pi k \gamma)^2+\gamma^2\sin(2\pi k)}},
\\
B_{c}=&\frac{2k}{2\pi k+\sin(2\pi k)}\left(\frac{1}{A_{c}}-2\pi\right),
\\
 A_{ch}=&\frac{\gamma^2\left[2\pi k+\sinh(2\pi k)\right]}
{k^3\pm k\sqrt{k^4-(2\pi k \gamma)^2+\gamma^2\sinh(2\pi k)}},
\\
B_{ch}=&\frac{2k}{2\pi k+\sinh(2\pi k)}\left(\frac{1}{A_{ch}}-2\pi\right),
\end{align}
and the frequency $k$ is real and the current $\gamma$ positive.
Note that $r_c$ and $r_{ch}$ solutions are related by $k\to i\,k$.
For a given $k$, $\gamma$ is limited by the square roots in $A_c$, $A_{ch}$ being real.
The phases, $\alpha$ and $\Omega$, are computed according to
Eqs.~(\ref{eq:phase}),~(\ref{eq:alpha}), and~(\ref{eq:omega}), respectively,
where now $k$ and $\gamma$ are taken as parameters,
and the chemical potentials read,
\begin{align}
\mu_{c}=&\frac{k^2}{2},
\\
\mu_{ch}=&-\frac{k^2}{2}.
\end{align}
The spectrum consists in a series of layered levels,
each one spanning all $\alpha$
and $\Omega\in[\frac{n}{2},\frac{n}{2}+\frac12]$, given that
$|k|\in[\frac{n}{2},\frac{n}{2}+\frac12]$, where $n=0,1,2,$ etc., see Fig.~\ref{fig:g0}.
Adding a perturbation to these solutions in the form of Eq.~(\ref{eq:pert})
must satisfy the same linear equations with $\mu\to\mu\pm w$,
as stated by Eqs.~(\ref{eq:bog1}) and~(\ref{eq:bog2}) with $g=0$. The solutions only satisfy the 
boundary conditions for real eigenvalues, and therefore the frequencies $w$ are not imaginary
and all the solutions stable.

\section*{Acknowledgments}
This work was supported by the Japan Ministry of Education, Culture, Sports, Science and Technology under the Grant number 15K05216.
We thank Muntsa Guilleumas, Bruno Juli\'a-D\'iaz and Ivan Morera
for fruitful discussions and a careful reading of the paper.
The article has also greatly benefited from ideas and comments
from the Atomtronics 2019 community.

\end{document}